\documentclass[aps,prc,onecolumn,nofootinbib]{revtex4}
\usepackage{epsf,epsfig,amssymb,amsmath}

\usepackage{color}

\newcommand\SKIP[1]{}
\newcommand{\be}{\begin{equation}}
\newcommand{\ee}{\end{equation}}
\newcommand{\bea}{\begin{eqnarray}}
\newcommand{\eea}{\end{eqnarray}}
\newcommand{\mybibitem}{\bibitem}

\renewcommand{\vec}[1]{{\bf #1}}

\newcommand{\gton}{\mathrel{\lower.9ex \hbox{$\stackrel{\displaystyle 
>}{\sim}$}}} 
\newcommand{\lton}{\mathrel{\lower.9ex \hbox{$\stackrel{\displaystyle 
<}{\sim}$}}}

\newcommand{\vx}{{\bf x}}

\newcommand{\vp}{{\bf p}}

\newcommand{\vv}{{\bf v}}
\newcommand{\vzero}{{\bf 0}}

\newcommand{\feq}{f^{\rm eq}}

\begin{document}

\title{Self-consistent conversion of a one-component bulk viscous fluid to particles}

\author{Denes Molnar}
\affiliation{Department of Physics and Astronomy, Purdue University, West Lafayette, IN 47907, USA}
\affiliation{Wigner Research Center for Physics, H-1525 Budapest, Hungary}

\date{\today}

\begin{abstract}
Comparison of heavy-ion experiments to fluid dynamics simulations
requires the conversion of the fluid to particles. 
Extending the approach in Ref.~\cite{Molnar:2014fva},
this work presents self-consistent bulk viscous corrections 
from kinetic theory for a one-component system with isotropic $2\to 2$
interactions. The phase space corrections are contrasted to 
the Grad ansatz and also to corrections obtained from 
the relaxation time approximation. In addition,
the bulk viscosity of the system is calculated and compared
with the Grad result, as well as the 
$\zeta \propto (1 - 3 c_s^2)^2 \eta$ relation between shear and 
bulk viscosity near the conformal limit.
The possible influence of various bulk correction choices on differential
elliptic flow $v_2(p_T)$ in heavy-ion collisions is also estimated.
\end{abstract}

\maketitle

\section{Introduction}

Much of our insight about ultrarelativistic heavy-ion reactions
comes from hydrodynamic 
modeling~\cite{Huovinen:2006jp,Gale:2013da,Heinz:2013th}.
An inevitable ingredient in these calculations,
if comparison to experiments is sought, 
is the conversion of the fluid to particles, 
called ``particlization'' \cite{Huovinen:2012is}.
As earlier works, e.g.,
\cite{Molnar:2011kx,Molnar:2014fva} have emphasized,
such a conversion is ambiguous 
for fluids with nonzero shear and/or bulk viscosity
because infinitely many choices for the particle phase space
densities can match the given hydrodynamic fields.
Analogous ambiguity is present even for ideal fluids~\cite{Takacs:2019ikb}, 
if one relaxes the implicit assumption of canonical statistics.

Phase space densities in particlization are often postulated 
using a convenient ansatz, for example, Grad's quadratic 
corrections~\cite{Song:2009rh,Monnai:2009ad}.
This, however,
ignores the microscopic dynamics that governs how the fluid departs from
local equilibrium. 
In contrast,
covariant kinetic theory provides a self-consistent theoretical framework
that relates dissipative corrections ($\delta f$) to scattering rates, 
as long as the fluid can be modeled as a gas mixture near freezeout.
The approach has been demonstrated in~\cite{Molnar:2014fva}
for shear viscous corrections,
where it was found that self-consistent corrections
have a weaker momentum dependence than the typically assumed quadratic 
terms. 
This difference affects identified particle harmonic
flow ($v_n$) and the extraction of shear viscosity 
as well\cite{Wolff:2016vcm} from data.
The self-consistent results for shear were also 
validated\cite{Damodaran:2020qxx}
in a comparison of several shear $\delta f$ models
against actual near-local-equilibrium evolution from covariant transport.

In this work, self-consistent bulk viscous phase space corrections are 
calculated from covariant transport
for a one-component system of particles
interacting with isotropic $2\to 2$ interactions.
Accurate results for the bulk viscosity of such a system are also obtained.

The self-consistent corrections are compared to the Grad 
ansatz (used, e.g., in the Hirano-Monnai bulk corrections 
\cite{Monnai:2009ad}), and the bulk $\delta f$ obtained 
from the relaxation time approach\cite{Bozek:2009dw, Dusling:2011fd}
(Dusling-Sch\"afer corrections).
These are both formulated in terms of additive contributions 
to the local equilibrium distribution, just like the self-consistent
results.
The comparison is not exhaustive -
particle distributions matching a set of hydrodynamic fields can also be 
postulated via rescaling temperature, chemical potential, 
{\em and} momenta in a thermal distribution (Pratt-Torrieri corrections~\cite{Pratt:2010jt}, or Tinti's approach \cite{Tinti:2015xwa}).
Nevertheless, valuable insights are gained into the systematic
errors made when bulk viscous fluids are particlized based on simplified
models.

\section{Bulk viscous phase space corrections}
\label{Sc:bulk}

The classification of dissipative corrections to the energy-momentum
tensor and conserved currents is well known
in the literature. Here we follow Sec. II of Ref.~\cite{Molnar:2014fva},
and focus on single-component systems. 

\subsection{Bulk pressure and $\delta f$}
\label{Sc:bulk_intro}

In the presence of bulk viscous corrections, the local energy-momentum
tensor of the fluid gets modified compared to its 
local equilibrium form by the presence of bulk pressure:
\be
T^{\mu\nu}_{\rm eq} = (e + p) u^\mu u^\nu - p g^{\mu\nu} \qquad \to \qquad 
T^{\mu\nu} = T^{\mu\nu}_{\rm eq} + \Pi (u^\mu u^\nu - g^{\mu\nu})\ .
\label{Tmunu_Pi}
\ee
Here, $u^\mu(x)$ is the local flow velocity, $e(x)$ is the local
energy density, $p(x)$ is the equilibrium pressure that is 
determined by the equation of state $p(e)$ of the fluid,
and $\Pi(x)$ is the bulk pressure that reflects spatially isotropic
local deviations from the equilibrium pressure 
in the local rest frame of the fluid (LR frame%
\footnote{In the LR frame, $u^\mu_{LR} = (1, \vzero)$.}%
).

By definition, the bulk correction does not contribute to the
local energy density, i.e., 
\be 
\delta e = u_\mu \delta T^{\mu\nu} u_\nu= 0 \ .
\label{delta_e}
\ee
Furthermore, if there are conserved charges, then the bulk correction 
also leaves the corresponding charge densities unchanged.
I.e., in the LR frame,
\be
\delta n_c = u_\mu \delta N^\mu_c = 0 \qquad (\forall\ {\rm conserved\ charge}\ c) \ ,
\label{delta_nc}
\ee
where $N^\mu_c$ is the charge current corresponding to conserved charge $c$.
The equation of state in this case is typically a
function of the LR-frame charge densities as well, i.e.,
$p(e, \{n_c\})$.

For small departures from equilibrium,
in the absence of transients (cf. Sec.~\ref{Sc:framework}), $\Pi$ is given
by the divergence%
\footnote{Minkowski scalar product of four-vectors $a$ and $b$ will
be denoted by the shorthand $(ab) \equiv a_\mu b^\mu$.}
of the flow velocity via the constitutive relation
\be
\Pi = - \zeta (\nabla u) \ ,
\label{Pi_with_zeta}
\ee
where $\zeta$ is the bulk viscosity, while 
$\nabla^\mu \equiv \partial^\mu - u^\mu (u \partial)$ 
denotes the component of the gradient orthogonal%
\footnote{Due to the constraint $u^2 = 1$, 
one could equivalently write
$\Pi = - \zeta (\partial u)$.}
to flow. However, this relation is insufficient for our purposes
because it does not give the phase space density of the particles
that make up the fluid.

In the presence of bulk pressure, particle phase space densities
get modified%
\footnote{Here, Boltzmann statistics is considered but extension to the
Bose or Fermi statistics is straightforward.}
in accordance with (\ref{Tmunu_Pi}):
\be
  \feq(x,\vp) \equiv \frac{g}{(2\pi)^3} e^{[\mu(x) - p_\nu u^\nu(x)]/T(x)} \qquad \to \qquad f(x,\vp) 
    = \feq(x, \vp) + \delta f(x, \vp) \ .
\ee
The general challenge for particlization is that infinitely many
choices for $\delta f$ can reproduce a given bulk correction to the 
energy-momentum tensor. 
That is because knowledge of the bulk pressure only constrains an 
integral of $\delta f$. 
In general,
\be
T^{\mu\nu}(x) = \int \frac{d^3 p}{E} p^\mu p^\nu f(x, \vp)  \ ,
\ee
where $E \equiv \sqrt{\vp^2 + m^2}$,
so 
\be
\delta T^{\mu\nu} = \Pi (u^\mu u^\nu - g^{\mu\nu})
= \int \frac{d^3 p}{E} p^\mu p^\nu \, \delta f \ ,
\label{Tmunu_df}
\ee
and projecting out the spatial diagonal elements in the LR frame
yields 
\be
\Pi \equiv - \frac{1}{3} (T_{LR})^i_{\ i} 
= \frac{1}{3} \int \frac{d^3 p}{E} \vp_{LR}^2 \, \delta f \ .
\label{Pi_from_Tmunu_p}
\ee
Using the constraint (\ref{delta_e}), one can also write
the above as
\be
\Pi = -\frac{1}{3} \delta T^{\mu}_{\ \mu}
 = -\frac{m^2}{3} \int \frac{d^3 p}{E} \delta f  \ .
\label{Pi_from_Tmunu_m}
\ee

It is customary in practice to
ignore the ambiguity and take an ansatz for $\delta f$.
In contrast, the self-consistent approach in Sec.~\ref{Sc:framework}
gives $\delta f$ as a solution to an integral equation.

\subsection{Grad ansatz for bulk corrections}
\label{Sc:grad}

A convenient ansatz~\cite{Monnai:2009ad} comes from
Grad's 14-moment approximation 
(Ch. VII.2 of \cite{deGroot}). For bulk corrections, the approximation is
\be
\delta f_{Grad} 
= \left(A + B \frac{E_{LR}}{T} + C \frac{E_{LR}^2}{T^2}\right) \feq \ ,
\label{Grad_df}
\ee
where $E_{LR} \equiv (p u)$ is the 
energy of the particle in the LR frame.
The prefactor in (\ref{Grad_df}) has a quadratic momentum dependence at high momenta.
Here $A$, $B$, and $C$ are dimensionless
constants that only depend on the particle mass to temperature
ratio $z \equiv m/T$, and are constrained by the conditions that
the bulk $\delta f$ contributes {\em neither} to the local energy density, 
{\em nor} to the local
particle density:
\bea
\delta e &=& \int \frac{d^3 p}{E} (p u)^2 \, \delta f_{Grad} = 0 \ ,
\label{Grad_delta_e}
\\
\delta n &=& \int \frac{d^3 p}{E} (p u) \, \delta f_{Grad} = 0 
\label{Grad_delta_n}
\eea
(the latter appears because 
particle density is conserved in $2\to 2$ scattering).
This means that only one of the coefficients is independent, 
and thus the full $\delta f_{Grad}$ 
is fixed by the bulk pressure $\Pi$.
The thermal integrals that appear in matching $A$, $B$, $C$, to $\Pi$ 
are collected in Appendix~\ref{App:Grad}.

The Grad ansatz is also commonly employed to provide analytic estimates of 
the bulk viscosity (cf. Sec. \ref{Sc:zeta}).

\subsection{Relaxation time approximation}
\label{Sc:rta}

Bulk viscous corrections may also be obtained in the relaxation
time approximation\cite{Bhatnagar:1954zz,Gyulassy:1997ib,Dusling:2011fd}.
As shown in \cite{Dusling:2011fd,Bozek:2009dw}, 
replacing the Boltzmann transport equation with the 
simplified linear equation
\be
p^\mu \partial_\mu f(x, \vp) 
= (pu) \frac{\feq(\vx, \vp) - f(\vx,\vp)}{\tau} \ ,
\ee
which is no longer an integral equation,
one obtains the bulk correction 
\be
\delta f_{DS} = const \times 
\left(\frac{p^2_{LR}}{3E_{LR} T} - c_s^2 \frac{E_{LR}}{T}\right) \, 
\feq  \ .
\label{DS_df}
\ee
Here, $\tau$ is a constant of time dimension that controls
the rate of scattering, and thus the dissipative corrections, while
\be
c_s^2 \equiv \frac{\partial p}{\partial e} 
= \frac{K_3(z)}{z K_2(z) + 3 K_3(z)}
\label{cs_squared}
\ee
is the speed of sound that is given explicitly in the last step for a gas in thermal
and chemical equilibrium (in terms of modified
Bessel functions, $K_n$). 

The correction (\ref{DS_df}) will be referred to here
as the Dusling-Sch\"afer (DS) form 
for brevity.
By construction, it satisfies the constraint (\ref{delta_e}),
just like Grad's ansatz, but its momentum dependence is quite
different.
At low momenta $\delta_{DS}/\feq$ is quadratic,
while at asymptotically high momenta it depends linearly on momentum.

\subsection{Self-consistent corrections from covariant transport theory}
\label{Sc:framework}

Self-consistent viscous corrections can be obtained from
covariant transport theory. The approach has been discussed in
depth in \cite{Molnar:2014fva} for shear viscous corrections 
from $2\to 2$ scattering. 
Here we apply it to calculate 
bulk viscous corrections for a single-component system.

\subsubsection{Covariant transport equation}
\label{Sc:cov_trans}

The starting point is the fully nonlinear Boltzmann transport equation
equation
\be
p^\mu \partial_\mu f(x,\vp) = S(x,\vp) + C[f](x, \vp) \ ,
\label{BTE}
\ee
where the source term $S$ encodes the initial conditions,
and the two-body collision term is
\be
C[f](x,\vp_1) 
\equiv \int\limits_2 \!\!\!\!\int\limits_3 \!\!\!\!\int\limits_4
\left(f_{3} f_{4} - f_{1} f_{2}\right)
\, \bar W_{12\to 34}  \, \delta^4(12 - 34)
\label{CollTerm}
\ee
with shorthands
$\int\limits_a \equiv \int d^3p_a / (2 E_a)$, 
$f_{a} \equiv f(x,\vp_a)$, and
$\delta^4(ab - cd) \equiv \delta^4(p_a + p_b - p_c - p_d)$.
The transition probability $\bar W_{12\to 34}$ 
for the $2\to 2$ process with momenta 
$p_1 + p_2 \to p_3 + p_4$
is invariant under interchange of incoming or outgoing particles,
\be
 \bar W_{12\to 34} \equiv \bar W_{21\to 34}
 \equiv \bar W_{12\to 43} 
 \equiv \bar W_{21\to 43} \ ,
\label{W_symmetry}
\ee
satisfies
detailed balance
\be
 \bar W_{34\to 12} 
      \equiv \bar W_{12\to 34} \ ,
\label{W_balance}
\ee
and is given by the corresponding unpolarized scattering matrix element
or differential cross section as 
\be
\bar W_{12\to 34} 
 = \frac{1}{16\pi^2} |\overline {{\cal M}_{12\to 34}}|^2
 \equiv \frac{4}{\pi} s p_{cm}^2 \frac{d\sigma_{12\to 34}}{dt}
\equiv 4s
      \frac{d\sigma_{12\to 34}}{d\Omega_{cm}} \ .
\label{W_with_sigma}
\ee
Here $s \equiv (p_1 + p_2)^2$ and $t \equiv (p_1 - p_3)^2$ 
are standard Mandelstam variables, while 
\be
p_{cm} \equiv \frac{\sqrt{(p_1 p_2)^2 - m^4}}{\sqrt{s}} = 
\frac{\sqrt{(p_3 p_4)^2 - m^4}}{\sqrt{s}}
\ee
is the magnitude of incoming (and, in our case, also outgoing) 
particle momenta 
in the center of mass frame of the microscopic two-body collision.

\subsubsection{Self-consistent bulk viscous corrections}
\label{Sc:selfcons_df}

For small departure from local equilibrium one can split the phase
space density into a local equilibrium part and a dissipative correction,
and linearize (\ref{BTE}) in $\delta f$:
\be
p^\mu \partial_\mu \feq + p^\mu \partial_\mu \delta f
=  \delta C[\feq,\delta f]
\label{linBTE}
\ee
with
\be
\delta C[\feq,\delta f](x,\vp_1) \equiv 
\int\limits_2 \!\!\!\!\int\limits_3 \!\!\!\!\int\limits_4
\left(\feq_{3} \delta f_{4} + \feq_4 \delta f_3 - \feq_{1} \delta f_{2} - \feq_2 \delta f_1 \right)
\, \bar W_{12\to 34}  \, \delta^4(12 - 34)
\ee
(the source term was dropped and space-time and momentum arguments are suppressed).
Typical systems quickly 
relax on microscopic scattering timescales to an asymptotic 
solution dictated
by gradients of the equilibrium distribution on the left hand side of 
(\ref{linBTE}), which is uniquely determined by 
the interactions in the system.
In this so-called Navier-Stokes regime the time derivative of 
$\delta f$ and also the spatial derivatives of $\delta f$
can both be neglected,
resulting in a linear integral equation to solve at each space-time point 
$x$. 
The same considerations appear in the standard calculation of 
transport coefficients in kinetic theory 
(see, e.g., Ch. VI of \cite{deGroot}, or \cite{AMYtrcoeffs}).

Bulk viscous correction are the response of the system to a nonuniform flow
velocity field with nonzero divergence but vanishing shear,
while temperature and chemical potentials are kept constant:
\be
(\partial u) \ne 0 \ , \qquad
\nabla^\mu u^\nu + \nabla^\nu u^\mu - \frac{2}{3} \Delta^{\mu\nu} (\partial u) = 0 \ , \qquad T = const \ , \qquad \mu_c = const \ ,
\ee 
where the tensor $\Delta^{\mu\nu} \equiv g^{\mu\nu} - u^\mu u^\nu$ projects
out vector components orthogonal to the flow velocity.
Under these conditions, the derivative on the LHS of (\ref{linBTE}) is
\be
(p\partial) \feq 
= - \frac{\feq}{3T} p_\alpha p_\beta \Delta^{\alpha\beta} (\partial u) \ .
\label{linBTE_bulk_source}
\ee
Upon decomposition into irreducible tensors in
momentum space according to $SO(3)$ 
representations in the LR frame (see App. A of \cite{Molnar:2014fva} 
and Refs.~\cite{deGroot,AMYtrcoeffs}),
the bulk viscous driving term (\ref{linBTE_bulk_source}) that is on the 
LHS of (\ref{linBTE}) corresponds to the scalar ($\ell = 0$) representation,
so the RHS of (\ref{linBTE}) must be in the same $\ell = 0$ representation.
This means that $\delta f$ itself must correspond to $\ell = 0$
because these representations are invariant subspaces of the 
linearized collision operator. 
Thus,
bulk viscous corrections are constrained to the form
\be
\delta f(x,\vp)
   = \chi(|\tilde\vp|) \frac{(\partial u)}{T} \feq(x,\vp)
\quad \quad {\rm with} \quad
\left.\frac{1}{T}\Delta^{\mu\nu} p_\nu\right|_{LR} \equiv (0, \tilde\vp) \ ,
\label{chi_def} 
\ee
where $\tilde\vp$ is the LR frame three-momentum normalized by temperature,
while $\chi$ is a real, dimensionless, 
scalar function of the rescaled momentum.
Substituting (\ref{chi_def})
into (\ref{linBTE}) yields, with the help of
\be
\feq_{3} \feq_{4} \delta^4(12-34) 
\equiv \feq_{1} \feq_{2} \delta^4(12-34) \ ,
\label{feq_convert}
\ee
the integral equation
\be
\frac{1}{3} \tilde p_1^2 \feq_{1} 
 = \frac{1}{T^2}
            \int\limits_2\!\!\!\!\int\limits_3\!\!\!\!\int\limits_4
         \feq_{1} \feq_{2} \, \bar W_{12\to 34}\,
         \delta^4(12-34)\, (  \chi_{3}  + \chi_{4}
                            - \chi_{1} - \chi_{2}) \ ,
\label{chi_eq}
\ee
with the shorthand 
\be
\chi_{a} \equiv \chi(|\tilde \vp_a|) \ .
\ee

It is straightforward to show with the help of
(\ref{W_symmetry}), (\ref{W_balance}) and (\ref{feq_convert}) 
that (\ref{chi_eq}) is equivalent to the extremization of the functional
\bea
Q[\chi] &=& -\frac{1}{3T^2} \int\limits_1 
                            \feq_{1}\, \chi_{1}\, \tilde p_1^2
        +\ \frac{1}{2T^4} 
                        \int\limits_1\!\!\!\!\int\limits_2\!\!\!\!
                        \int\limits_3\!\!\!\!\int\limits_4
           \feq_{1} \feq_{2} \, \bar W_{12\to 34}\,
           \delta^4(12-34)\,
          (  \chi_{3} + \chi_{4} 
           - \chi_{1} - \chi_{2}) \chi_{1}
\nonumber \\
&\equiv& B 
         + (Q_{31} + Q_{41} - Q_{11} - Q_{21}) 
     \ ,
\label{Qdef}
\eea
i.e., (\ref{chi_eq}) is reproduced by the usual variational procedure imposing
$\delta Q[\chi] = 0 + {\cal O}(\delta\chi^2)$. This allows one to
estimate $\chi$ variationally using a finite basis $\{\Psi_{n}\}$ as
\be
\chi(|\tilde\vp|)
    = \sum\limits_n c_{n} \Psi_{n}(|\tilde\vp|)
\label{chi_expand}
\ee
and finding optimal coefficients $\{c_{n}\}$ that maximize $Q$.
If the basis is complete, the limit $n\to \infty$ reproduces the 
exact solution. Numerical evaluation of $Q$ is discussed in 
Appendix~\ref{App:Qintegrals}.

Unlike shear corrections, 
the general expansion (\ref{chi_expand}) for bulk corrections
does not {\em automatically} satisfy the matching conditions 
(\ref{delta_e}) and (\ref{delta_nc}).
This is because the
densities involved are also scalars ($\ell = 0$) under rotations of momenta 
in the LR
frame, just like the bulk correction%
\footnote{
Whereas shear viscous corrections correspond to angular momentum 
$\ell = 2$, and thus contribute to neither the energy density nor 
any charge densities.
}.
Therefore, $Q[\chi]$ must be extremized under the constraints 
(\ref{delta_e}) and (\ref{delta_nc}),
i.e.,
\bea
&&\int \frac{d^3p}{E} (p u)^2 \,\chi \feq = 0 \ ,
\nonumber\\
&& \int \frac{d^3p}{E} (p u) \,\chi \feq = 0 \ .
\label{dQ_constraints}
\eea
These can be straightforwardly implemented using Lagrange multipliers, 
i.e., via extremizing the extended functional
\be
Q'[\chi] = Q[\chi] 
+ \alpha \int \frac{d^3p}{E} (p u)^2 \,\chi \feq 
+ \beta \int \frac{d^3p}{E} (p u) \,\chi \feq 
\label{Qprime_chi}
\ee
with respect to both $\chi$, and the parameters $\alpha$ and $\beta$.
Although the constraints affect the solution for $\chi$,
both $Q'$ and $Q$ evaluate to the same value at the solution.

\subsubsection{Relation to bulk viscosity}
\label{Sc:bulk_viscosity}

The form of the self-consistent bulk correction (\ref{chi_def})
implies the constitutive relation (\ref{Pi_with_zeta}), therefore,
one can calculate the bulk viscosity from the solution $\chi$.
With the help of (\ref{Pi_from_Tmunu_p}) and (\ref{Qdef}) one has
\be
\zeta = - \frac{\Pi}{(\partial u)} 
= - 2T \int_1 \frac{\tilde p_1^2}{3} \chi_1 \feq_1 = 2 T^3 B[\chi],
\ee
where the last equality applies when $Q[\chi]$ is maximized.
The maximal value of the quadratic functional $Q$, on the other hand, is
one-half of its linear term, so $Q_{max} = B/2$.
Thus, the bulk viscosity is given by the maximum of the functional 
$Q[\chi]$ as
\be
\zeta = 4 Q_{max} T^3 \ .
\label{zeta_from_Q}
\ee

Any ansatz for $\chi$
can be used to estimate the bulk viscosity from below
as $\zeta > 4T^3 Q[\chi]$,
as long as $\chi$ satisfies the constraints (\ref{dQ_constraints}).
Bulk viscosity results from the Grad ansatz (\ref{Grad_df}) are discussed 
in Sec.~\ref{Sc:zeta}.

\section{Results}

Self-consistent bulk corrections and bulk viscosities have been 
computed, numerically, for a wide range of particle masses and temperatures
$10^{-2} \le m/T \le 70$. A variety of different basis sets
have been explored, such as 
\bea
\phi_n^{(1)}(x) &=& x^n \ ,  \quad  n = 0, 1, 2, ... \ ,
\\
\phi_n^{(2)}(x) &=& x^{n/2} \ , \quad n = 0, 1, 2, ... \ ,
\\
\phi_n^{(3)}(x) &=&  x^{n/4} \ , \quad n = 0, 1, 2, ... \ ,
\\
\phi_n^{(4)}(x) &=& (x^2 + z^2)^{n/2} \ , \quad n = 0, 1, 2, ... \ .
\eea
Sets 1-3 correspond to integer and fractional 
powers of the rescaled LR-frame momentum,
while basis set 4 to integer powers of the rescaled LR-frame energy.
Fastest convergence was observed for basis set 2 but
all basis choices converged eventually to the same answer.

The Grad ansatz corresponds to set 4 with three basis functions,
$n = 0$, 1, and 2.

\subsection{Bulk viscous corrections for one-component system}
\label{Sc:chi}

\begin{figure}[h]
\leavevmode
\begin{center}
\epsfysize=6cm
\epsfbox{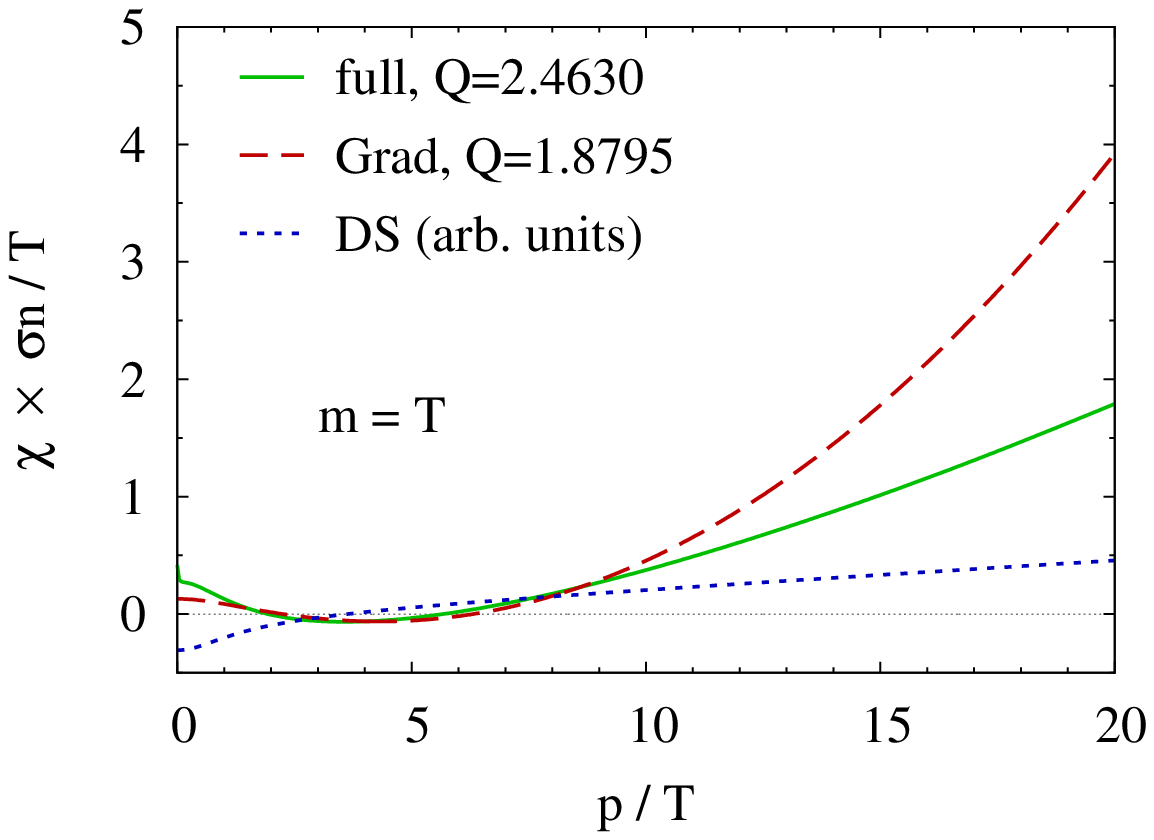}
\hskip 0.5cm
\epsfysize=6cm
\epsfbox{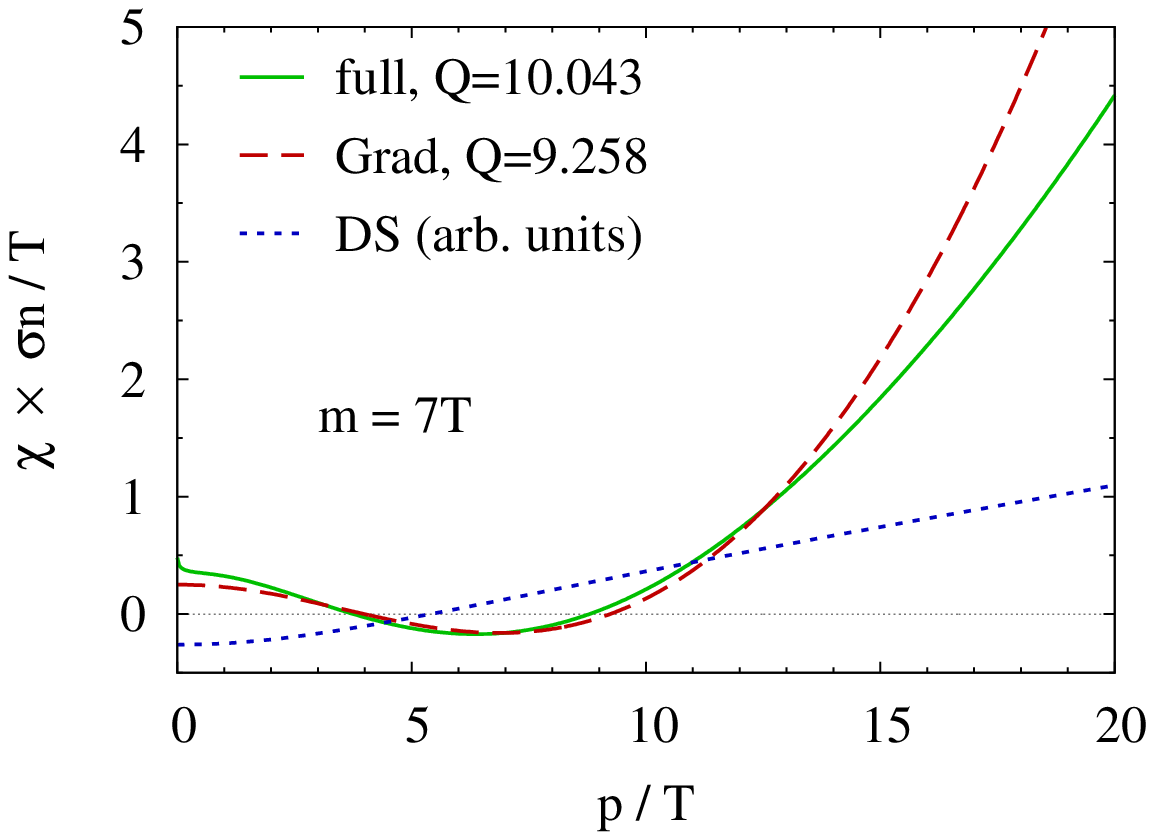}
\caption{Self-consistent bulk viscous corrections (solid green lines) 
vs $p/T$ contrasted with
the Grad approximation (dashed red lines),
for $m/T$ = 1 (left panel) and 7 (right panel)
calculated with constant, isotropic $2\to 2$ cross section. 
Corrections are normalized by the dimensionless mean 
free path
$T \lambda \equiv T/n\sigma$. The value $Q$ in the legends indicates the 
corresponding variational result for the bulk viscosity,
$\zeta = 10^{-3} \times Q T/\sigma$.
For comparison, the shape of the Dusling-Sch\"afer bulk viscous correction 
is also shown (dotted blue lines).
}
\label{Fig:1}
\end{center}
\end{figure}

Figure~\ref{Fig:1} shows results for the bulk viscous correction 
$\chi$ as a function of rescaled LR-frame momentum, for two different 
mass to temperature ratios $m/T = 1$ (left panel) and 7 (right panel). 
Given typical freeze out temperatures of $T\sim 120-140$~MeV 
in heavy-ion collisions, 
the former is relevant for pions at freezeout, while the latter for 
protons. Self-consistent corrections are proportional%
\footnote{The bulk correction $\delta f$, and thus $\chi$ as well,
encodes per-particle viscous effects. The bulk pressure is the
sum of such contributions from all particles, therefore,
density dependence drops out from the bulk viscosity 
- only $T/\sigma$ remains (cf. (\ref{zeta_Grad})).}
to the 
dimensionless mean free path 
\be 
T \lambda_{MFP} = \frac{T}{n\sigma}  \ , 
\ee 
so those are shown with dimensionless mean free path scaled out.
The same factor
has been scaled out from the Grad results as well because those correspond
to a self-consistent calculation performed in a limited variational basis.

Unlike 
self-consistent shear viscous corrections~\cite{Molnar:2014fva}, which 
are monotonic in momentum and exhibit approximate power-law dependence 
with an exponent close to 3/2, self-consistent bulk corrections 
have a more complicated structure.
This is not surprising 
because certain positively-weighted integrals (\ref{dQ_constraints}) of 
the bulk $\chi$ vanish, therefore, $\chi$ must switch sign
at least once.
In fact, the self-consistent result (solid green) switches sign twice
- it is positive at low momenta and later at high momenta as well 
but goes negative in between. 
The Grad ansatz (dashed red), which is quite similar,
though not identical, exhibits the same behavior.
For comparison, the Dusling-Sch\"afer form (dotted blue),
which comes from the relaxation time approximation,
switches sign only once.

Due to its restricted variational basis,
the Grad result noticeably overpredicts the correction at high momenta,
especially for the lower $m/T = 1$ (pions). This is also reflected in
the lower maximum value achieved for $Q[\chi]$ (shown in the legends
multiplied by a factor $10^3 \times 2 \sigma T^2$), which translates 
into an underestimated bulk 
viscosity from the Grad ansatz.

\subsection{Mass dependence of bulk viscosity}
\label{Sc:zeta}

For a system of point-like particles with 
$2 \to 2$ interactions, 
bulk viscosity vanishes on general grounds in both the massless $m \to 0$ (UR)
and nonrelativistic $m \to \infty$ (NR) limits. 
In the massless limit, this is because 
the energy-momentum tensor is traceless.
So, in local equilibrium
\be
\left( T_{\rm eq}\right)^{\mu}_{\ \mu} = e - 3p = 0 \ ,
\label{Tr_Tmunu_eq}
\ee 
i.e., the equilibrium pressure is $p = e/3$,
while in the presence of any bulk correction (\ref{Tmunu_Pi}), 
\be
T^{\mu}_{\ \mu} = e - 3(p + \Pi) = 0 \ ,
\ee
which is inconsistent with (\ref{Tr_Tmunu_eq}) unless $\Pi = 0$.

In the nonrelativistic limit,
the reason why bulk viscosity vanishes lies in the matching conditions
(\ref{delta_e}) and (\ref{delta_nc}),
namely, that the bulk correction leaves the local comoving energy density
and particle density unchanged. This implies that the LR-frame
kinetic energy density is also unchanged
\be
\delta (e - m n) = \delta e - m \delta n = \int \frac{d^3 p}{E} (E_{LR}^2 - m E_{LR}) \delta f = 0 \ .
\label{ekin_NR}
\ee
For large $z$, the factor in the integrand can be written as
\be
E_{LR}^2 - m E_{LR} = p_{LR}^2 + m^2 - m^2 \sqrt{1 + \frac{p_{LR}^2}{m^2}}
= \frac{p_{LR}^2}{2} 
  \left[1 + {\cal O}\!\left(\frac{p_{LR}^2}{m^2}\right)\right] \ ,
\ee
so (\ref{ekin_NR}) becomes
\be
0 = \int \frac{d^3 p}{E} \frac{p_{LR}^2}{2} 
    \left[1 + {\cal O}\!\left(\frac{p_{LR}^2}{m^2}\right)\right] \, \delta f\ .
\label{ekin_NR_2}
\ee
Comparison to (\ref{Pi_from_Tmunu_p}) tells that the first term is $3\Pi / 2$,
and in the NR limit the correction term 
(coming from the relativistic ${\cal O}(p^4/m^3)$ correction to the energy) 
is dropped straight away, so $\Pi = 0$.
If in doubt, the correction term can be estimated based on the observation 
that in the NR limit
\be 
\feq_{NR}(\vp) = \frac{n}{(2\pi m T)^{3/2}} e^{-p^2 / 2mT}
\ee
is Gaussian in momentum, 
and the thermal average of each power of $p^2$ gives a factor proportional
to $mT$.
With the reasonable assumption that at high momenta $\chi$ 
is dominated by a power-law,
the additional factor of $p_{LR}^2/m^2$ in the
integrand contributes a factor on the order of $(mT)/m^2$ to the integral.
Thus, (\ref{ekin_NR_2}) gives
\be
0 = \frac{3\Pi}{2} \left[1 + {\cal O}\left(\frac{T}{m}\right)\right] \ ,
\ee
which in the $m\to \infty$ limit indeed
leads to $\Pi \to 0$ and, therefore, $\zeta \to 0$.

The analytic result in the Grad approximation (\cite{deGroot}, App. XI) 
matches these expectations:
\be
\zeta_{Grad} = \frac{z^2 K_2^2(z) 
                    [(5 - 3\gamma)\hat h - 3\gamma]^2}
                   {16[2 K_2(2z) + z K_3(2z)]}
        \frac{T}{\sigma} \ ,
\label{zeta_Grad}
\ee
where 
\be
\hat h \equiv \frac{h}{n} = \frac{z K_3(z)}{K_2(z)}  
\ee
is the enthalpy per particle, while 
\be
\gamma \equiv \frac{c_p}{c_v} = 1 + \frac{1}{z^2 + 5 \hat h - \hat h^2 - 1}
\ee
is the ratio of 
constant pressure and constant volume heat capacities.
Near the UR and NR limits,
\be
\zeta_{Grad}(z\ll 1) \approx \frac{z^4}{72} \frac{T}{\sigma}\ , \quad
\zeta_{Grad}(z\gg 1) \approx  \frac{25\sqrt{\pi}}{64 z^{3/2}} \frac{T}{\sigma}\ ,
\label{zeta_Grad_limits}
\ee
which both vanish in their respective limits. 

\begin{figure}[h]
\leavevmode
\begin{center}
\epsfysize=6cm
\epsfbox{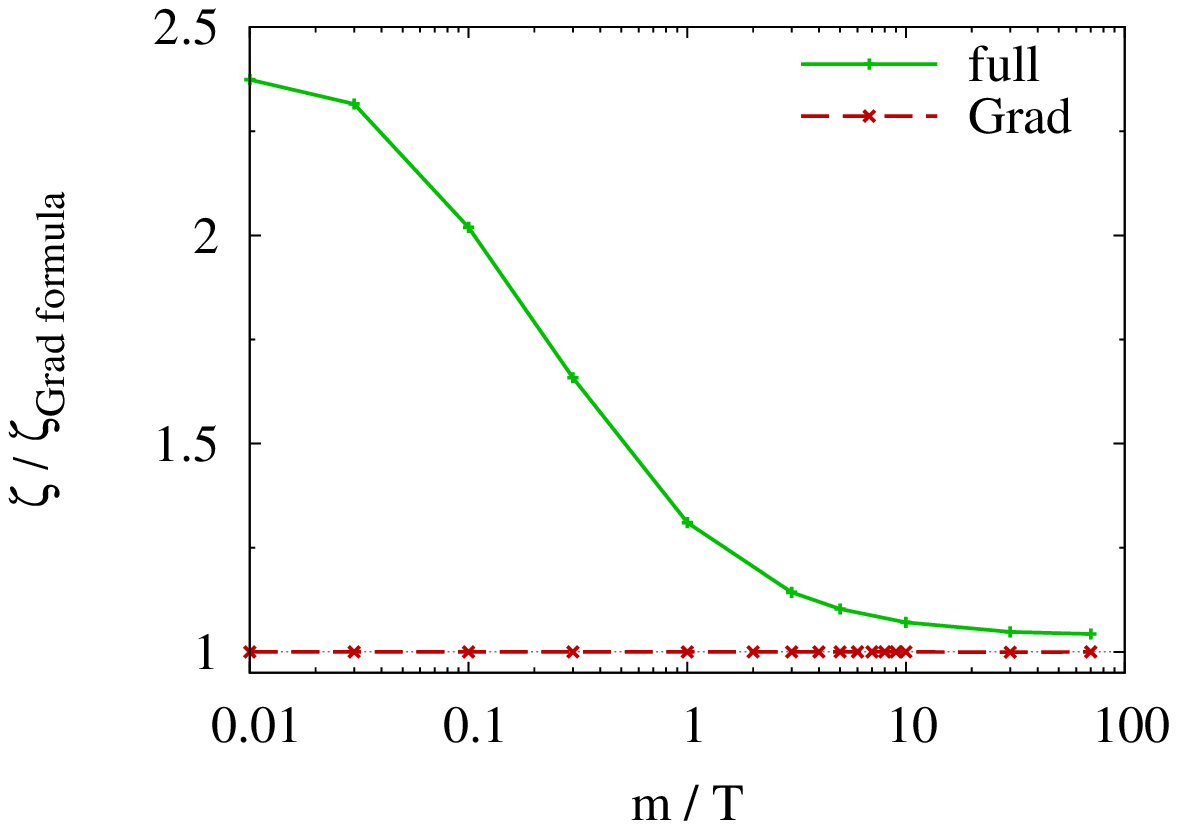}
\epsfysize=6cm
\epsfbox{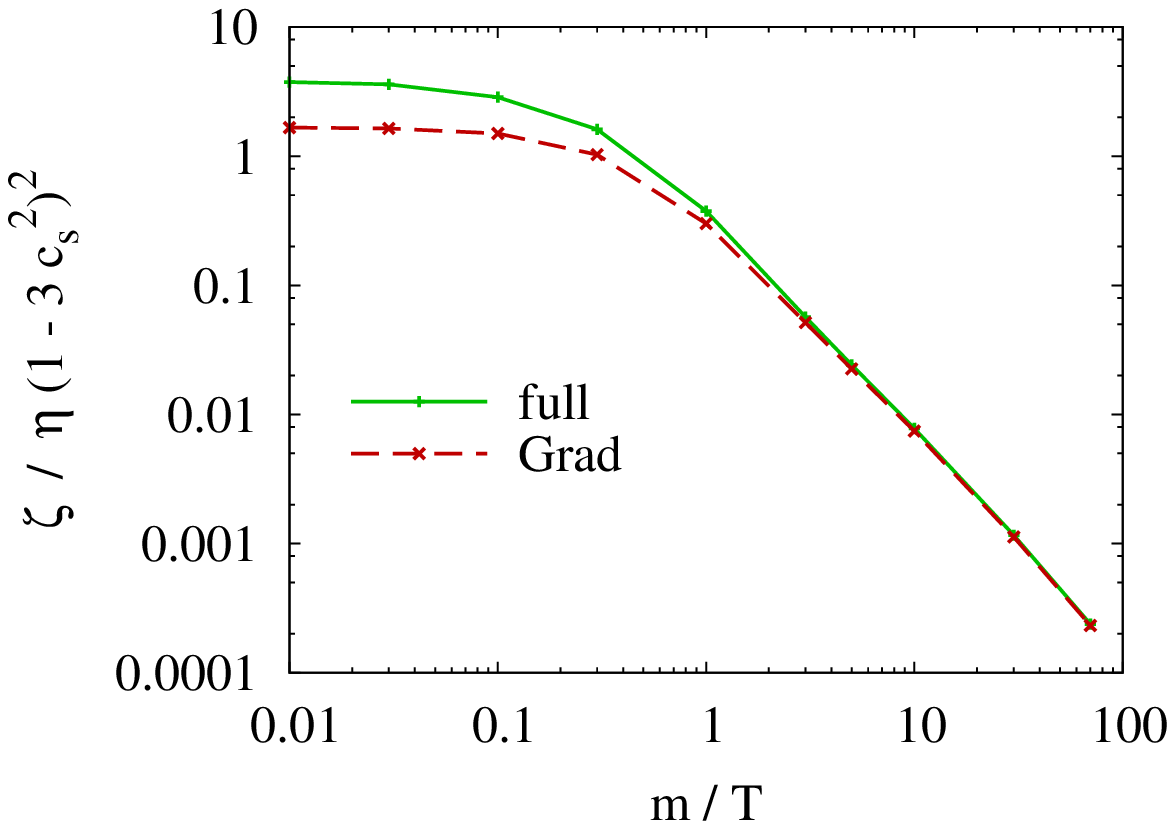}
\caption{{\em Left panel:} bulk viscosity  {\em relative} to the
analytic Grad result as a function of normalized particle mass $m/T$
for a one-component
system with constant, isotropic $2\to 2$ cross sections. 
The full self-consistent solution is shown in solid green. 
To demonstrate the accuracy of the calculation,
the Grad result was also obtained numerically using the respective
limited variational basis (red crosses). 
{\em Right panel:} comparison to the near-conformal 
$\zeta \propto (1 - 3 c_s^2)^2 \eta$
relation\cite{Weinberg:1971mx} between bulk and shear viscosities, 
as a function of $m/T$.
Results from self-consistent bulk {\em and} shear corrections
(solid green line) are compared to those obtained using the Grad approximation
for both bulk and shear (red dotted line). 
Shear viscosities were calculated based on~\cite{Molnar:2014fva}.
}
\label{Fig:2}
\end{center}
\end{figure}

Figure~\ref{Fig:2} shows the numerically calculated bulk viscosity 
obtained with the self-consistent bulk viscous corrections.
The left panel shows $\zeta$ as a 
function of the mass to temperature ratio $m/T$,
normalized to the analytic Grad result 
(\ref{zeta_Grad}). 
As expected, the Grad approximation generally underpredicts the
bulk viscosity. Near the NR limit, i.e., for large masses, the 
error is modest, less than about 10\% for $z > 10$.
On the other hand, in the opposite (small mass) limit,
the Grad ansatz underpredicts the bulk viscosity by a progressively
larger factor that exceeds 2 for $z < 0.1$.
To illustrate the accuracy of the calculations, 
numerical results are also shown from the limited variational basis used 
in the Grad approximation (red dashed line), with excellent agreement 
with the analytic result in the full mass range $0.01 \le m/T \le 70$ 
studied here.

The right panel of Fig.~\ref{Fig:2} compares the shear and bulk viscosities
for a one-component system with energy-independent, isotropic $2\to 2$
interactions, to the observation $\zeta \propto (1 - 3 c_s^2)^2 \eta$
made by Weinberg\cite{Weinberg:1971mx} near the conformal limit.
The same relationship was 
also studied in \cite{Dusling:2011fd}. 
As can be seen (solid green line), the proportionality indeed holds within good
accuracy for $m/T \lton 0.03$, in fact,
\be
 \zeta \approx 3.75 (1 - 3 c_s^2)^2 \eta \ .
\ee
The numerical coefficient 3.75 changes to about 1.67 if one compares the approximate
shear and bulk viscosities from the Grad approach (dashed red line),
in good agreement with the analytic expression%
\footnote{The analytic result is a combination of
(\ref{zeta_Grad_limits}), $\eta_{Grad} = 6T/5\sigma$ for $z=0$,
and $c_s^2 \approx 1/3 - z^2/36$ from (\ref{cs_squared}).
}
\be
\zeta_{Grad} \approx \frac{5}{3} (1 - 3 c_s^2)^2 \eta_{Grad} \ .
\ee

\subsection{Effect on differential elliptic flow}
\label{Sc:v2}

Finally, we give an estimate for how much the choice of bulk viscous 
correction model affects differential elliptic flow in ultrarelativistic 
nucleus-nucleus collisions. To that end we employ a bulk viscous 
generalization of the simple four-source model in 
Ref.~\cite{Huovinen:2001cy}, which models a snapshot of the system via 
four fireballs 
moving back-to-back along the $x$ and $y$ directions with velocities 
$\pm v_x$ and $\pm v_y$, respectively, in the transverse plane of the 
collision. The volumes and temperatures of all four sources are set to be 
the same but with $v_x > v_y > 0$, which leads to a positive 
elliptic flow as a function of transverse momentum (except for heavy particles
at low momenta). Bulk viscous 
corrections are introduced here analogously to the shear viscous
generalization of the model in \cite{Takacs:2019ikb} 
via assuming that in the rest frame of each fireball viscous parameters are 
the same.
Bulk pressure is a Lorentz scalar, 
so we set the bulk pressure over energy density ratio to 
the same value $\Pi / e = -0.001$ in all four fireballs
(in heavy-ion collisions, $(\partial u)$ is typically positive, so 
$\Pi$ is typically negative).

The combined momentum distribution of particles from all four fireballs,
is then
\be
f^{(4s)} = f_{(v_x, 0)} + f_{(-v_x, 0)} + f_{(0, v_y)} + f_{(0, -v_y)} \ ,
\ee
where subscripts denote the velocity of the sources in the transverse plane.
For a single source,
the momentum distribution of particles at midrapidity,
i.e., at momenta
\be
\vp = (p_x, p_y, 0) \equiv p_T (\cos \varphi, \sin \varphi, 0) \ ,
\ee
is
\be
f_{\vec v} (p_T, \varphi) = const \times \left[1 + a \frac{\Pi}{e}\,\chi(\frac{p_{LR}}{T})\right] e^{-(pu)/T} \ ,
\ee
where
\bea
 (pu) &=& \frac{1}{\sqrt{1 - \vv^2}} \left[\sqrt{p_T^2 + m^2} - p_T (v_x \cos \varphi + v_y \sin \varphi)\right] \ ,
\\
 p_{LR} &=& \sqrt{(pu)^2 - m^2} \ ,
\eea
and the scaling factor $a$ is set such that the bulk pressure of the
fireball reproduces the $\Pi/e$ ratio with the given bulk
viscous correction model ($a = 0$ gives back the thermal
sources in the original four-source model).
Differential elliptic flow can now be calculated, numerically,
directly from its definition as
\be
v_2(p_T) \equiv \frac{\int_0^{2\pi} d\varphi \, \cos 2\varphi \, f^{(4s)}(p_T,\varphi)}
                     {\int_0^{2\pi} d\varphi \, f^{(4s)}(p_T, \varphi)} \ .
\ee

\begin{figure}[h]
\leavevmode
\begin{center} 
\epsfysize=6cm
\epsfbox{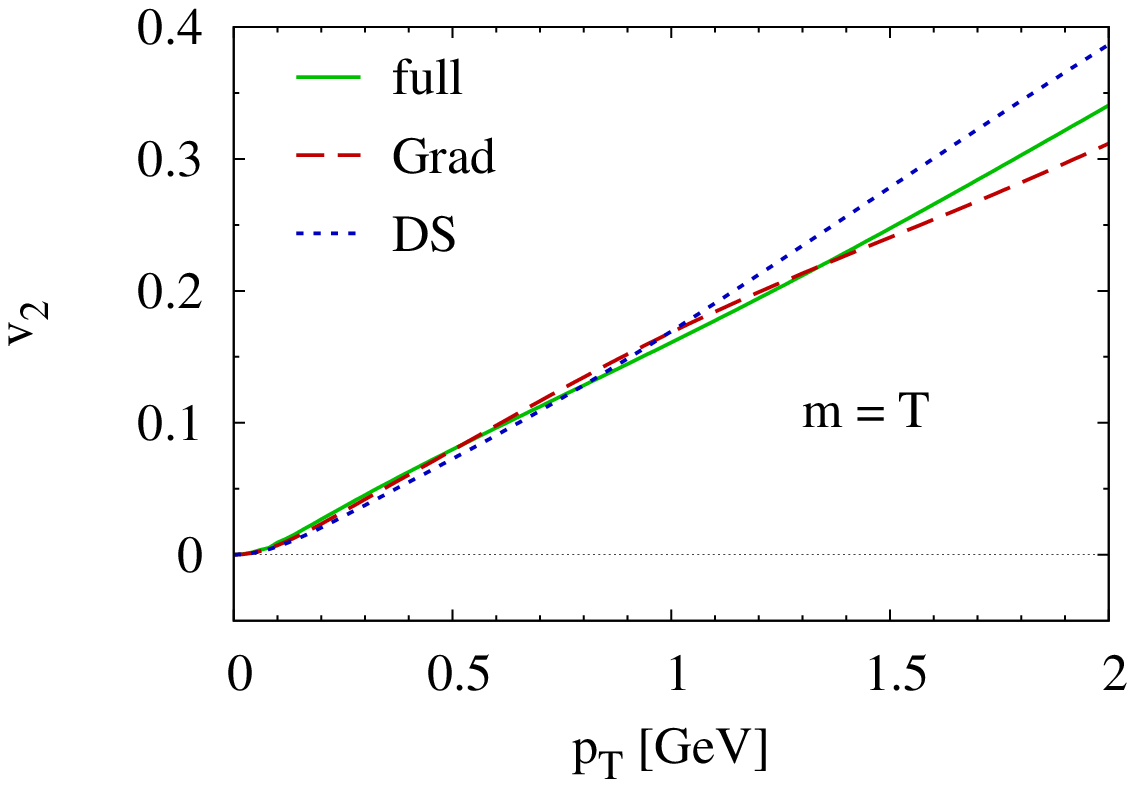}
\hskip 0.5cm
\epsfysize=6cm 
\epsfbox{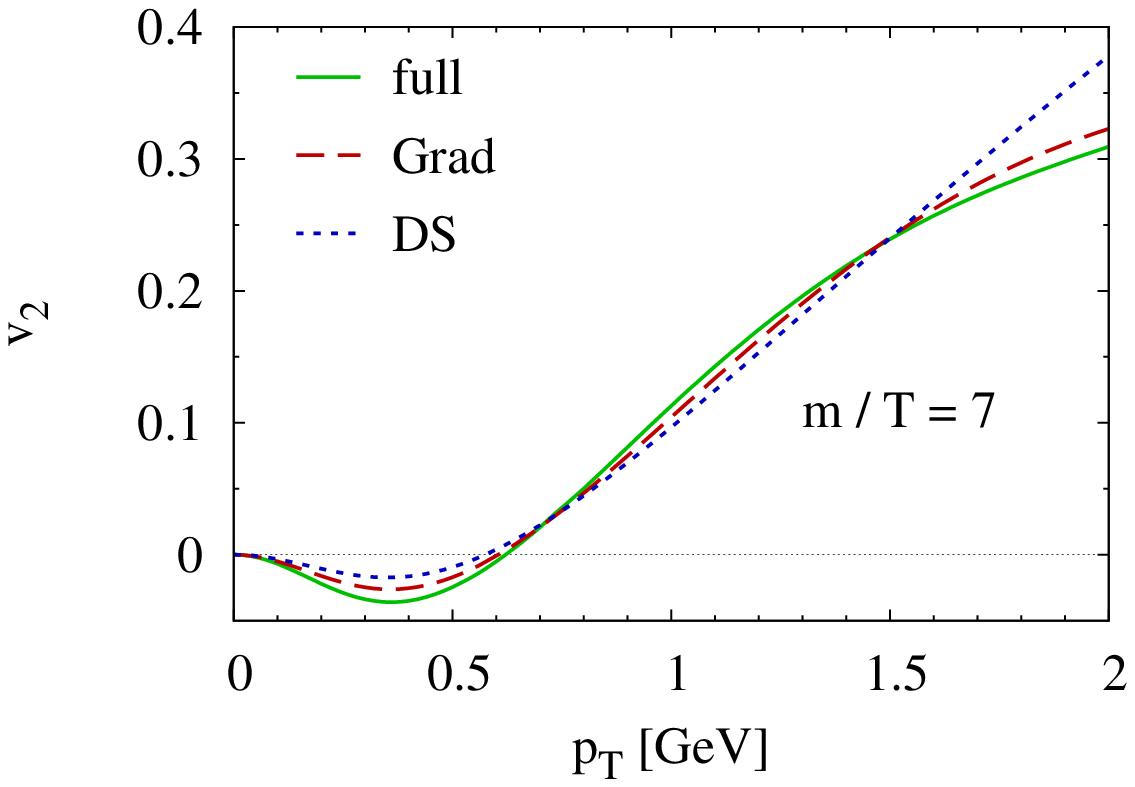} 

\caption{Differential elliptic flow $v_2(p_T)$ vs transverse momentum 
from a generalization of the four-source model in 
Refs.~\cite{Huovinen:2001cy,Takacs:2019ikb} for pions ($m/T = 1$, left 
panel) and protons ($m/T = 7$, right panel). 
Three bulk viscous corrections models are compared: 
i) self-consistent bulk corrections from $2\to 2$ kinetic theory (solid 
green lines), 
ii) the Grad approximation (red dashed lines), and
iii) corrections based on the relaxation time 
approximation~\cite{Dusling:2011fd} (dotted blue lines).
The model parameters were $T = 0.14$~GeV, $v_x = 0.6$, $v_y = 0.5$,
$\Pi/e = -0.001$.
}
\label{Fig:3}
\end{center}
\end{figure}

Figure~\ref{Fig:3} shows the results for differential
elliptic flow for parameters $T=0.14$~GeV, $v_x = 0.6$, $v_y = 0.5$
that approximate mid-central Au+Au collisions at top RHIC energy 
$\sqrt{s_{NN}} = 200A$~GeV, for $m/T = 1$ (left panel) and $7$ (right panel),
appropriate for pions and protons at freezeout, respectively.
In both cases, rather small differences are seen between the
full self-consistent results (solid green lines) and the Grad
ansatz (red dashed line); the largest difference is an about
10\% smaller $v_2$ from the Grad approach at the highest $p_T= 2$~GeV
plotted.
In contrast, bulk corrections from the relaxation time
approximation (dotted blue line) give significantly larger $v_2$
than the other two $\delta f$ models at high $p_T \gton 1.5$~GeV.
In fact, this latter model gives almost identical $v_2$ to that from 
pure thermal sources (not shown). This means that the bulk viscous correction
to pion $v_2$ is about $50$\% percent larger (more negative)
with the Grad ansatz, in this calculation, 
than the bulk correction from the self-consistent approach.

A reliable assessment of bulk viscous effects on heavy-ion
observables will, of course, have to take into account interactions
in the multicomponent hadron gas (i.e., mixtures). 
As has been shown
for the case of shear~\cite{Molnar:2014fva}, this leads to intricate
inter-species dependences. For example, 
even when one restricts the calculation to the Grad ansatz for each species,
viscous corrections acquire species-dependent scaling factors 
that cannot be obtained from calculations for single-component systems.

\section{Conclusions}
\label{Sc:concl}

In this work self-consistent bulk viscous phase space corrections ($\delta f$)
are calculated from covariant kinetic theory,
for one-component systems with isotropic, energy-independent
$2\to 2$ interactions. Compared to self-consistent shear viscous
corrections\cite{Molnar:2014fva}, 
which show an approximate power-law momentum dependence, 
bulk corrections exhibit more complicated behavior and
change sign twice as a function of momentum.

The corrections are contrasted
with the Grad ansatz, which postulates a quadratic polynomial
of the particle energy in the frame comoving with the fluid,
and also with bulk viscous corrections based on the relaxation time 
approximation~\cite{Dusling:2011fd} (DS model). 
While the Grad bulk $\delta f$ is quite
similar to the self-consistent results, except at high momenta, the
relaxation time approximation leads to bulk viscous corrections with 
markedly different functional shapes.

The above findings are reflected in estimates for
bulk viscous corrections to differential elliptic flow
in A+A reactions at RHIC energies, obtained from a simple semi-analytic
four-source model motivated by Refs.~\cite{Huovinen:2001cy} and 
\cite{Takacs:2019ikb}. The self-consistent and Grad bulk $\delta f$ models
give nearly identical proton $v_2(p_T)$ 
in the $0 < p_T < 2$~GeV window studied, while for pions at high 
$1.5 < p_T <2$~GeV the Grad approach overestimates bulk viscous effects
by up to $50$\%. Under the same conditions, the bulk $\delta f$ from
the DS approach generates negligible bulk viscous corrections to $v_2$.

The bulk viscosity of the system is also calculated and compared to 
known analytic formulas (\cite{deGroot}, App. XI) derived in the Grad approximation.
For large masses $m > 10 T$, 
the Grad result is accurate to within $10$\%,
however, for small $m < 0.1T$ 
it underestimates the bulk viscosity by more than a factor of 2.
We also find that near the massless limit the Weinberg relation\cite{Weinberg:1971mx}
between bulk and shear viscosities 
$\zeta = const \times (1 - c_s^2)^2 \eta$ 
does hold with a coefficient of about 3.75, 
which is roughly a factor of two larger 
than the coefficient of 5/3 in the Grad approximation.

While the above results are intriguing,
they are limited to single-component systems with
$2\to 2$ interactions. It would be interesting
to follow up this investigation, in the future,
with a calculation of self-consistent bulk viscous corrections 
for hadronic mixtures.

\acknowledgments 
We thank the hospitality of the 
Wigner Research Center for Physics (Budapest, Hungary) 
where parts of this work were completed. 
Computing
resources managed by RCAC/Purdue and the GPU Laboratory of Wigner RCP 
are gratefully acknowledged.
This work was supported by the U.S. 
Department of Energy, Office of Science, under Award No. DE-SC0016524,
and by the Hungarian National Research, Development
and Innovation Office (NKFIH) under contract number OTKA K135515
and THOR Cost Action CA15213.

\appendix

\section{Grad correction matching}
\label{App:Grad}

This section lists the thermal integrals that appear while
matching the Grad correction (\ref{Grad_df}) to a given bulk pressure.
For a local local equilibrium distribution
\be
\feq = \frac{g}{(2\pi)^3} e^{[\mu - (p u)]/T} \ ,
\ee
the constraints (\ref{Grad_delta_e}) and (\ref{Grad_delta_n})
can be written as
\bea
j_0 A + j_1 B + j_2 C &=& 0 \ ,
\label{Grad_constraint_1}
\\
j_1 A + j_2 B + j_3 C &=& 0 \ ,
\label{Grad_constraint_2}
\eea
where
\be
j_n \equiv \int_0^{\infty} dx x^2 y^n e^{-y}  
\propto \int \frac{d^3 p}{E} (p u)^{n + 1} f_{\rm eq}
\ee
with $y \equiv \sqrt{x^2 + z^2}$ and $z \equiv m/T$. 
After switching to $y$ as the integration
variable, the integrals yield modified Bessel functions of the
second kind:
\bea
j_0 &=& z^2 K_2(z) \ ,
\\
j_1 &=& z^3 K_1(z) + 3 z^2 K_2(z) \ ,
\\
j_2 &=& z^4 K_2(z) + 3 z^3 K_3(z) \ ,
\\
j_3 &=& 2 z^4 K_2(z) + (15 + z^2) z^3 K_3(z) \ .
\eea

Matching to the bulk pressure gives the third constraint.
Specifically, from (\ref{Pi_from_Tmunu_m}):
\bea
\Pi &=& \frac{(-z^2)}{3} \frac{g}{2\pi^2} T^4 e^{\mu/T} 
     \int_0^\infty dx x^2 \frac{1}{y} (A + B y + C y^2) e^{-y} 
\nonumber \\
&=& \frac{g}{2\pi^2} T^4 e^{\mu/T} \frac{(-z^2)}{3} (j_{-1} A + j_0 B + j_1 C) \ ,
\eea
where $j_{-1} = z K_1(z)$.
Substitution of
the local equilibrium energy density
\be
e \equiv \int \frac{d^3p}{E} (p u)^2 f_{\rm eq} = \frac{g}{2\pi^2} T^4 e^{\mu/T} j_1
\ee
leads to
\be
\frac{\Pi}{e} = -\frac{z^2}{3} \left(\frac{j_{-1}}{j_1} A + \frac{j_0}{j_1} B + C\right) \ .
\label{Grad_constraint_3}
\ee
The linear system (\ref{Grad_constraint_1}), (\ref{Grad_constraint_2}),
and (\ref{Grad_constraint_3}) can now be solved in a straightforward
manner for $A$, $B$, and $C$ in terms of $\Pi/e$.

\section{Calculation of momentum integrals in {\boldmath $Q[\chi]$}}
\label{App:Qintegrals}

The integrals in (\ref{Qdef}) can be evaluated using analogous
steps to the shear viscous case discussed in App. B of \cite{Molnar:2014fva}
(see Sec. \ref{Sc:framework} for the definitions of the usual shorthands).
The only difference is that, for bulk viscous corrections, 
the contraction
\be
P_a\cdot P_b \equiv P_a^{\mu\nu} P_{b,\mu\nu} 
  = (\tilde \vp_a \tilde\vp_b)^2
                       - \frac{1}{3}|\tilde \vp_a|^2 |\tilde \vp_b|^2
\ee
does not arise, and the source $B$ comes from the divergence
of the flow instead of its shear.

For a one-component system the flavor indices play no role 
and, therefore, will be dropped 
(we have $ij \to k\ell$ scattering with $i=j=k=\ell$).
Thus,
\be
B = -\frac{2\pi}{3T^4} \int\limits_{m}^\infty 
       dE_1 \, p_1^3 \feq_{1} \chi(\frac{p_1}{T})
 \ .
\ee
Next, $Q_{11}$ and $Q_{21}$
are given by Eqns. (B4) and (B5) in Ref.~\cite{Molnar:2014fva},
but without the extra factor $P_1 \cdot P_1 = 2p_1^4/3T^4$ in $Q_{11}$ and
$P_2 \cdot P_1 = p_1 ^2 p_2 ^2 (t_{12}^2 - 1/3)/T^4$ in $Q_{21}$:
\bea
Q_{11} &=& \frac{2 \pi^2}{T^4} 
\int\limits_{m}^\infty dE_1\, p_1\, \feq_{1}\,
\chi_{1}^2
\int\limits_{m}^\infty dE_2\,  p_2\, \feq_{2}
  \int\limits_{-1}^{1} {dt_{12}} \, F(s)\, \sigma_{TOT}(s)  \ ,
\label{Q11_integration}
\\
Q_{21} &=& \frac{2\pi^2}{T^4}
\int\limits_{m}^\infty dE_1\, p_1\,\feq_{1}\,
\chi_{1}
\int\limits_{m}^\infty dE_2\, p_2\,\feq_{2}\,
\chi_{2}
  \int\limits_{-1}^1 dt_{12} \,
 F(s)\, \sigma_{TOT}(s) \ ,
\label{Q21_integration}
\eea
where
\be
F(s) \equiv p_{cm} \sqrt{s} 
    = \frac{1}{2}\sqrt{s (s - 4 m^2)} \ ,
\label{Fs}
\ee
and $\sigma_{TOT}$ is the total cross section.
The integration variables $E_1$ and $E_2$ correspond to the
LR-frame energies of the incoming particles in $2\to 2$ scattering,
while $t_{12}$ is the cosine of the angle between
incoming momenta in the LR frame.

Interchange symmetry (\ref{W_symmetry})
with $3\leftrightarrow 4$ implies $Q_{41} = Q_{31}$,
so the last remaining contribution to discuss is $Q_{31}$.
For isotropic cross section, it can be reduced
to four integrals using the steps in
App.~B.2 of Ref.~\cite{Molnar:2014fva}.
In fact, in the bulk viscous case the averaging over the c.m. frame
angle $\phi_3$ 
is trivial because the integrand does
not depend on $\phi_3$ at all.
Therefore, for general energy-dependent cross sections we have
\be
\int\limits_1\!\!\!\!
\int\limits_2\!\! \int d\Omega_3 \,(...)
\ \to\  \frac{4\pi \cdot 2\pi \cdot 2\pi}{4} 
\int\limits_{m}^\infty dE_1\, p_1 
\int\limits_{m}^\infty dE_2\, p_2
\int\limits_{-1}^1 dt_{12} 
\int\limits_{-1}^1 dt_3 (...) \ ,
\ee
i.e.,
\be
Q_{31} = \frac{\pi^2}{T^4}
\int\limits_{m}^\infty dE_1\, p_1\,\feq_{1}\, \chi(\frac{p_1}{T})
\int\limits_{m}^\infty dE_2\, p_2\,\feq_{2}
\int\limits_{-1}^1 dt_{12} F(s) \sigma_{TOT}(s) 
\int\limits_{-1}^1 dt_3 \, \chi(\frac{p_3}{T})\ ,
\label{Q31_integration}
\ee
where 
\be 
 p_3 = |\vp_3| = \sqrt{E_3^2 - m^2} 
     = \sqrt{(\gamma_3 E_T + \beta_3 p_T t_3)^2 - m^2}
\ee
 with
\bea
&&
\beta_3 \equiv \frac{p_{cm}}{\sqrt{s}} \quad, \qquad
\gamma_3 \equiv \frac{E_{3,cm}}{\sqrt{s}} 
= \sqrt{\beta_3^2 + \frac{m^2}{s}} \quad, \qquad
\\
&& E_T \equiv E_1 + E_2 \quad , \qquad 
p_T \equiv |\vp_1 + \vp_2| = \sqrt{p_1^2 + p_2^2 + 2 p_1 p_2 t_{12}} \ .
\eea
The variable $t_3$ is the cosine of the angle of deflection
in the microscopic scattering (in the center-of-mass frame).


\end{document}